\documentclass[prl,twocolumn,groupedaddress,showpacs]{revtex4}
\usepackage{amsmath,amsfonts,amssymb}
\usepackage{graphicx}
\newcommand{\upp}{\hspace{-2 pt}\uparrow}
\newcommand{\downn}{\hspace{-2 pt}\downarrow}

\begin{document}

\bibliographystyle{apsrev}

\title{Controlling a mesoscopic spin environment by quantum bit manipulation}
\author{J.M. Taylor$^1$, A. Imamoglu$^2$, and M.D. Lukin$^1$}
\address{
$^1$ Department of Physics, Harvard University, Cambridge,
Massachusetts 02138, USA \\
$^2$ Institute of Quantum Electronics, ETH-H\"onggerberg, HPT G
12, Z\"urich, Switzerland}

\begin{abstract}
We present a unified description of cooling and  manipulation of a mesoscopic bath of nuclear spins via coupling to 
a single quantum system of electronic spin (quantum bit). 
We show that a bath cooled by the quantum bit
rapidly saturates.
Although the resulting saturated states of the spin bath (``dark states'') generally have low degrees of polarization and
purity, their symmetry properties make them  a valuable resource for the coherent manipulation of quantum bits. 
Specifically, we demonstrate that the dark states of nuclear ensembles can be used to coherently control the system-bath interaction
and to provide a robust, long-lived quantum memory for qubit states.
\end{abstract}
\maketitle

An intriguing challenge for modern science and technology is the coherent manipulation of 
quantum systems coupled to realistic environments. Interest in these problems is in part due to fundamental 
aspects of quantum control and decoherence, but this research has also been stimulated by recent developments in 
quantum information science~\cite{neilsen}. Although over past decades much progress has been made in the controlled manipulation
of isolated atomic and optical systems~\cite{monroe}, the complex environment of a solid-state system makes it significantly more
challenging to achieve a similar degree of control.

This Letter  demonstrates that a single quantum system (qubit) can be used  to prepare and 
control a mesoscopic environment, 
turning the bath into a useful resource.  
We consider a system consisting of a single electronic spin in a semiconductor quantum dot  interacting with  a
mesoscopic bath of nuclear spins within the confined volume. 
Recently it has been shown that cooling the spin bath to high
values of polarization and purity greatly reduces the associated decoherence~\cite{khaetskii}. Furthermore, due to the bath's intrinsic memory, it can be used as a  long-lived quantum memory for qubits and for quantum state engineering of collective nuclear states~\cite{taylor03}.   However, achieving a high degree of nuclear polarization in a quantum dot remains a major experimental challenge.  Most ideas under exploration use
the hyperfine contact interaction to couple polarized electron spins to the bath.
Some work {\em in situ}, 
  using either spin-polarized
currents~\cite{eto02} or optical pumping~\cite{gammon,imamoglu03}.
Other techniques use a different geometry for cooling, such as quantum Hall edge state tunneling near a quantum point contact~\cite{dixon97}.

We focus on {\em in situ} manipulation, when the qubit degrees of freedom are themselves used to cool nuclei.  
We show that such a qubit-based cooling process rapidly saturates resulting in non-thermal states of the nuclear bath with low polarization and purity.  However, the symmetry properties of such  ``dark states''  allow for 
complete control of the qubit's interaction with the environment. We illustrate this by showing that the mesoscopic bath prepared in saturated states can be used to provide a long-lived quantum memory for qubit states.
A combination of adiabatic passage techniques and spin-echo results in near unity storage  fidelity even for the bath 
with vanishingly small polarization. 
Before proceeding we also note that the idea of using  qubit states to cool the environment
is now widely applied in atomic systems such as trapped ions~\cite{wineland} or microwave cavity QED~\cite{haroche}.

A simple Hamiltonian can describe a single electron spin confined 
in a quantum dot interacting with an applied external magnetic 
field $B_0$ and with $N$ spin-$I_0$ surrounding nuclei via the 
hyperfine contact interaction:
\begin{equation}
\hat{H} = g^* \mu_B B_0 \hat{S}_z + g_n \mu_n B_0 \sum_k \hat{I}^k_z + a \sum_k \alpha_k \hat{\vec{S}} \cdot \hat{\vec{I}}^k .
\end{equation}
The hyperfine interaction is split into a field aligned (Overhauser) component $\hat{V}_{zz} = a \hat{A}_z \hat{S}_z$ and a Jaynes-Cummings type component, $\hat{H}_{JC} = a/2 (\hat{A}_+ \hat{S}_- + \hat{A}_- \hat{S}_+)$, where $\hat{\vec{A}} = \sum_k \alpha_k \hat{\vec{I}}^k$, $\alpha_k = N v_0 | \psi( r_k )|^2$ is the weight of the electron wavefunction at the $k$th lattice site, and $a$ is the per nucleus hyperfine interaction constant.  

Nuclear degrees of freedom are cooled by cycling spin-polarization 
through the quantum dot.
A spin-down electron is injected from leads connected to a polarized reservoir or by means of optical excitation.  It interacts for some short time $\tau$, and then is ejected / recombined.  Each iteration can cool the bath by flipping a nuclear spin through $\hat{H}_{JC}$.  If the energy difference of the injected electron spin and the flipped electron spin, $\langle \Delta \rangle = (g^* \mu_B - g_n \mu_n) B_0  + a \langle \hat{A}_z - 1 \rangle$  is large compared to the inverse time of interaction, $\tau^{-1}$, energy conservation considerations block the spin-flip process.
However, changing the applied field to maintain $\langle \Delta \rangle \tau \ll 1$ allows cooling to continue efficiently~\cite{imamoglu03}.

Regardless of the exact details of the process, cooling {\it will} 
saturate.  The system is driven into a statistical mixture of 
``dark states'' $|\mathcal{D}\rangle$, defined by~\cite{imamoglu03}
\begin{equation}
{\hat A}_- |\mathcal{D}\rangle = 0 .
\end{equation}
To cool past this point of saturation, either dark states must couple to other states of the bath or the geometric coupling coefficients $\alpha_k$ must change.  When 
these mechanisms are slow compared to the cooling rate, an appropriate mixture of dark states well approximates the steady state of the bath.

\begin{figure}
\includegraphics[width=2.5in]{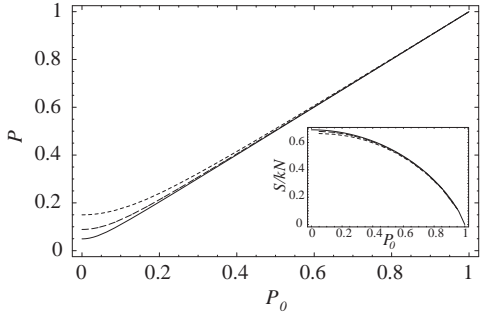}
\caption{
Saturated state polarization $P$ versus initial thermal polarization $P_0$ for $N$=100(dotted), $300$(dashed) and $10^3$ (solid).  Inset shows entropy per spin. \label{f:polpol}
}
\end{figure}

The homogeneous case illustrates the essential features of cooling.
With $\alpha_k = 1$, we can rewrite
$\hat{\vec{A}}$ as a collective nuclear angular momentum vector $\hat{\vec{J}}$, and correspondingly, $\hat{J}^2$ becomes a conserved quantity. The Dicke basis, characterized by total (nuclear) angular momentum $J$ ($0,1/2\le J\le N/2$), its projection into the z axis $m_J$, and
a quantum number associated with the permutation group $\beta$, is then  appropriate~\cite{arechi72}. 
The operator $\hat{J}_-$ changes neither $J$ nor $\beta$, but nuclei in a state $|J, m_J, \beta \rangle$ are cooled to the state with lowest $m_J$ (dark state) $|J,-J,\beta\rangle$.  For an initial thermal bath of nuclei with polarization $P_0$, the corresponding steady state solution is found by summing over $-J \le m_J \le J$.  Tracing over $\beta$, we find 
\begin{eqnarray}
\hat{\rho}_{ss} &=& \sum_J \rho_n(J) |J, -J \rangle \langle J, -J | =  (2 \cosh(\kappa / 2))^{-N} \nonumber \\
&\times&
 \sum_{J} D(J) \frac{\sinh[\kappa/2 (2 J +1)]}{\sinh[\kappa/2]} | J, -J \rangle \langle J, -J | ,
\label{e:steadystate}
\end{eqnarray}
with $\kappa = 2 \tanh^{-1}(P_0)$. $D(J)$ denotes the number of $\beta$ quantum numbers allowed for a given $J$ and is independent of $m_J$.  In the case of spin-1/2 nuclei, $D(J) ={N\choose N/2 - J} - {N \choose N/2-J-1}$.    The resulting nuclear polarization $P$ and Von Neumann entropy associated with the ``cooled'' ensemble are shown
 in Figure~\ref{f:polpol} as a function of initial thermal polarization  $P_0$. 
The differences shown in Fig.~\ref{f:polpol} between the thermal and saturated baths becomes negligible for large $N$, but the dynamics of the two baths differ dramatically.  In essence, even though the purity and polarization are low, the symmetry properties of the dark states restricts evolution of the combined electronic-nuclear system, analogous to a two-level system.

We illustrate the reversible nature of the coupling between dark states and the single spin by showing how a quantum state can be stored into collective nuclear states.  An arbitrary qubit state $|\Psi\rangle = u |\upp \rangle + v |\downn \rangle$ will be mapped into the bath states. 
With just a pure state $|J, -J \rangle$, spin-down is decoupled entirely, while spin-up couples nuclei to 
the collective state $|J,-J + 1\rangle$ with a Rabi frequency $\Omega_J = a \sqrt{2 |J|}$. As
\begin{equation}
\hat{H}_{JC}^2 |\upp \rangle | J, -J \rangle  =  \case{a^2}{2} J | \upp \rangle | J, -J \rangle \nonumber ,
\end{equation}
the motion is given by the cyclic dynamics of a two-level system.  
Near resonance ($|\langle \Delta^2 \rangle| \lesssim |\Omega_J|^2$),
the qubit will oscillate fully between electronic and nuclear states.  For high polarization $P$ this is in direct analogy to the case discussed in Ref.~4.  However, for low $P$, all states of the mixture must be in resonance; the range of $J$ with significant probability goes as the width of the binomial distribution, $\sqrt{N}$, while the width of the resonance, given by $\langle \Omega_J \rangle/a$, goes as $\sqrt{P N}$.  In this regime, the resonance is much narrower than the range of populated $J$ states.

This problem can be solved with adiabatic passage.  By sweeping the detuning from far negative to far positive, the system passes through the series of avoided crossings and for each $J$
\begin{equation}
(u | \upp \rangle + v | \downn \rangle) | J, -J \rangle \rightarrow
| \downn \rangle (u e^{i \phi_J} | J, -J+1 \rangle + v | J, -J \rangle. \nonumber
\end{equation}
Adiabatic passage is not sensitive to the exact value of the coupling constant (Rabi frequency) between individual pairs of levels, and is robust provided the sweep rate of the detuning is sufficiently slow.

In general, the relative phase $\phi_J$ accumulated depends on the details of the detuning sweep.  Sweeping the detuning back reverses storage, but the final state, $( u e^{2 i \phi_J} | \upp \rangle + v | \downn \rangle) | J, -J \rangle$, has an additional non-trivial phase which reduces the final off-diagonal matrix element of 
electronic spin density matrix: $\rho_{\uparrow \downarrow} = u v^* [\sum_J \rho_n(J) e^{2 i \phi_J}$].
Spin-echo avoids this strong dephasing by exactly compensating the adiabatically acquired phase~\cite{symmodes}.
  An example sequence is presented in Table~\ref{t:adiabatic}. The two waiting segments should be symmetric, to compensate for other arbitrary $J$-dependent phases~\cite{decohstuff}.
  Thus a mixture of saturated states can be used as an ideal quantum memory.

\begin{table}
        \begin{center}
\begin{tabular}{||c|c|c||} \hline
${\rm state} $ & $ {\rm process} $ & $ \Delta(t)  $ \\ \hline
$(u | \upp \rangle + v | \downn \rangle) | J, -J \rangle $ & $ {\rm start} $ & $ \Delta_i $ \\
$ | \downn \rangle ( u e^{i \phi_J} | J, -J +1 \rangle + v | J, -J \rangle) $ & $ {\rm store} $ & $	\rightarrow \Delta_f $ \\
 & $ {\rm wait} $ & $ \Delta_f $ \\
$(u e^{2 i \phi_J} | \upp \rangle + v | \downn \rangle) | J, -J \rangle $ & $ {\rm retrieve} $ & $  \Delta_i \leftarrow $ \\
$(i v | \upp \rangle + u e^{2 i \phi_J} | \downn \rangle) | J, -J \rangle $ & $ \pi-{\rm pulse} $ & $ \Delta_i $ \\
$| \downn \rangle ( i v e^{i \phi_J} | J, -J +1 \rangle + u e^{2 i \phi_J}  | J, -J \rangle) $ & $ 
	{\rm store} $ & $   	\rightarrow \Delta_f $ \\
	& $ {\rm wait} $ & $ \Delta_f $ \\
$(i v e^{2 i \phi_J} | \upp \rangle + u e^{2 i \phi_J} | \downn \rangle) | J, -J \rangle $ & $ {\rm retrieve} $ & $  \Delta_i \leftarrow $ \\
$e^{2 i \phi_J} (u | \upp \rangle - v | \downn \rangle) | J, -J \rangle $ & $ \pi-{\rm pulse} $ & $ \Delta_i $ \\ \hline
\end{tabular}
\caption{Adiabatic transfer with ESR spin echo. 
\label{t:adiabatic}}
\end{center}
\end{table}

In practice, the electron-spin decoherence rate $\gamma$ limits the minimum speed of the adiabatic sweep, the induced error scaling as $p_{\gamma} \simeq \gamma T$  with $T$ as the characteristic duration of the storage procedure.  For a saturated ensemble state, $T \simeq 4 \langle \Omega_{J} \rangle/\xi$, where $\xi = \langle \dot{\Delta} \rangle$ is the rate of change of the detuning.  
 Assuming a  tangent-like pulse shape~\cite{pulsecite} we find that the non-adiabatic probability of flip is
given by
\begin{equation}
p_{na} \simeq \xi^2 / 32 \langle \Omega_{ J } \rangle^4 = 1/2 T^2 \langle \Omega_J \rangle^2 .
\end{equation}
The total error probability is then $p_{tot} = p_{\gamma} + p_{na}$.
Minimizing this for $T$ gives $T_{min} = (\gamma \langle \Omega_{ J }\rangle^2)^{-1/3}$ and 
\begin{equation}
p_{tot,min} = 3/2 (\gamma / \langle \Omega_{ J }\rangle)^{2/3}.
\end{equation}

The saturated state lifetime, the storage lifetime, and the maximum polarization are limited by nuclear spin dephasing.  Spin diffusion due to dipolar nuclear coupling is the dominant term for this dephasing in GaAs and is on the order of $6 \times 10^4~{\rm s}^{-1}$~\cite{paget}; this rate also provides an estimate for the rate of heating from the proximal thermal spins.  Active correction pulse sequences such as WHH-4 can lead to sub-Hz decoherence rates and lowered spin diffusion~\cite{mehring}.  Finally, we note that these results generalize
 to higher spin by using the appropriate multinomial form of $D(J)$~\cite{taylorTBP}.

We now extend these results to realistic inhomogeneous coupling between electrons and
nuclei by developing a one-to-one mapping between the explicit homogeneous Dicke basis and its inhomogeneous equivalent.  Dicke states of the form $| J, -J, \beta \rangle$ in the individual spin basis are written
\begin{equation}
|J, -J=n-N/2, \beta \rangle = \sum_{\{j\}_{n}} c_{J,\beta}( \{ j \} )    | \{ j \} \rangle 
\end{equation}
where the set $\{ j \}_n$ labels $n$ spins that are pointing up; the rest point down.  As $\hat{J}_- | J, -J, \beta \rangle = 0$, the $c$-numbers $c_{J,\beta}(\{ j \})$ must satisfy
\begin{equation}
\sum_{l \notin \{ i \}_{n-1}} c_{N/2-n,\beta}(\{ i \} + l) = 0 \label{e:c_prop}
\end{equation}  
for all $\{ i \}_{n-1}$.
Furthermore, $\hat{J}_-$ is invariant under permutation so there exists a representation for dark states where every individual spin configuration is equally probable, {\em i.e.} $|c_{J,\beta}(\{i \})|^2 = |c_{J,\beta}(\{ j \})|^2 = {N \choose N/2-J}^{-1}$.  
Using this explicit representation for homogeneous dark states, we construct a mapping to the more general inhomogeneous case $(\hat{A}_- | \mathcal{D}(n,\beta) \rangle = 0)$.  For each dark state $|J, -J=n-N/2, \beta \rangle$, its inhomogeneous counterpart is 
\begin{equation}
|\mathcal{D}(n,\beta) \rangle = \mathcal{N}_{0,0}^{-1/2} \sum_{\{ j \}_n} \left( \prod_{ k \in \{ j \}} \frac{1}{\alpha_k} \right) c_{N/2-n,\beta}(\{ j \}) | \{ j \} \rangle, \label{e:darkstate}
\end{equation}
as can be checked by direct calculation. The exact form of the normalization constant  $\mathcal{N}_{0,0} $ is defined below. 

To quantify inhomogeneous effects, first we note that $\hat{A}_- \hat{A}_+$ maps $|\mathcal{D}(n,\beta) \rangle$ into an orthogonal state $ |\mathcal{O}(n,\beta) \rangle$, $\hat{A}_- \hat{A}_+ |\mathcal{D}(n,\beta) \rangle = |\Omega_n|^2 |\mathcal{D}(n,\beta) \rangle + | \chi_n|^2 |\mathcal{O}(n,\beta) \rangle$ with 
\begin{eqnarray}
\Omega_n & = & a \sqrt{\langle \mathcal{D}(n,\beta) | \hat{A}_- \hat{A}_+ |\mathcal{D}(n,\beta) \rangle} , \nonumber \\
\chi_n & = & a  \left(\langle \mathcal{D}(n,\beta) | \hat{A}_- \hat{A}_+ \hat{A}_- \hat{A}_+ |\mathcal{D}(n,\beta) \rangle - |\Omega_n|^4 \right)^{1/4} \nonumber .
\end{eqnarray}
Non-zero $\chi_n$ indicates that an inhomogeneous equivalent of $\hat{J}^2$ is not conserved under inhomogeneous raising and lowering operators.
Second, inhomogeneous dark states are also not eigenstates
of $\hat{A}_{z}$, {\em i.e.} $\hat{V}_{zz} |\mathcal{D}(n,\beta)\rangle = \delta_{n} |\mathcal{D}(n,\beta) \rangle + \omega_n |\mathcal{B}(n,\beta) \rangle$, where 
\begin{equation}
\omega_n  = \sqrt{\langle \mathcal{D}(n,\beta) | \hat{V}_{zz}^2 |\mathcal{D}(n, \beta) \rangle - \langle \mathcal{D}(n,\beta) | \hat{V}_{zz} |\mathcal{D}(n, \beta) \rangle^2}. \nonumber 
\end{equation}

If the symmetry breaking terms $(\chi_n, \omega_n)$ are small 
relative to $\Omega_n$, cooling will proceed in a manner similar 
to the homogeneous case.  The final state density matrix should 
then be of the type $\hat{\rho} = \sum_{n,\beta} \rho(n) 
|\mathcal{D}(n,\beta) \rangle \langle \mathcal{D}(n,\beta) |$.  
When $\chi_n$ and $\omega_n$ are small we can use Eqn.~\ref{e:steadystate} as an estimate for $\rho(n)$.  As before, 
cooling proceeds quickly to the point of saturation, then slows down to a rate governed by the inhomogeneous transfer of dark states into other states.

Adiabatic transfer of a quantum state follows the prescription for the 
homogeneous case.  However, the symmetry 
breaking terms lead to additional errors.  When $\chi_n \ll \Omega_n$, the rate of transfer is given by $\Omega_n$ but the final state is an admixture of $|\mathcal{D}(n,\beta) \rangle$ and $|\mathcal{O}(n,\beta) \rangle$, leading to an error of order $\chi_n^2 / \Omega_n^2$. 
The error from $\omega_n$ we estimate in the worst case by considering it as an incoherent loss mechanism.  The effective decoherence rate become $\gamma_{eff,n} = \sqrt{\omega_n^2 + \gamma^2}$, the spin-decoherence rate optimization used for the homogeneous case holds, and the resulting probability of error for the full sequence goes as $3/2 (\gamma_{eff,n} / \Omega_n)^{2/3}$.  Combining these, the total probability of error goes as 
\begin{equation}
p_{tot} = 1 - \sum_n \left\{ \left(\frac{\Omega_n^2}{\Omega_n^2 + \chi_n^2} \right)^4 \left[1 - 3 \left(\frac{\gamma_{eff,n}}{ \Omega_n} \right)^{2/3}\right] \rho(n) \right\} . \nonumber
\end{equation}

\begin{figure}
\includegraphics[width=2.5in]{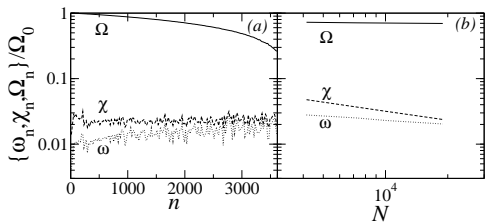}
\caption{
{\em (a)} $\Omega_n$ (solid), $\omega_n$ (dotted) and $\chi_n$ (dashed)  versus number of flipped spins $n$ with $N=7280$.  Values are averaged over every 20 $n$ values and scaled by $\Omega_0 = A/\sqrt{N}$.  {\em (b)} Average values of $\Omega_n/\Omega_0,~\omega_n/\Omega_n$, and $\chi_n/\Omega_n$ vs. $N$.
\label{f:inhomog_fid}}
\end{figure}

We now consider adiabatic transfer errors numerically.
The explicit form of the inhomogeneous dark states (\ref{e:darkstate}) allows us to express the relevant parameters as functions of the geometric coupling constants, $\alpha_k$:
\begin{eqnarray}
\Omega_n^2 / a^2& = & \sum_k \alpha_k^{2} - 2 \mathcal{N}_{2,1}(n)/\mathcal{N}_{0,0}(n) , \\
\chi_n^4 / a^4 & = & \frac{4 \mathcal{N}_{2,2}(n)}{\mathcal{N}_{0,0}(n)}   - \left(\frac{2\mathcal{N}_{2,1}(n)}{\mathcal{N}_{0,0}(n)}\right)^2 , \\
\omega_n^2/a^2 & = &  \frac{\mathcal{N}_{1,2}(n)}{ \mathcal{N}_{0,0}(n)} - \left(\frac{\mathcal{N}_{1,1}(n)}{ \mathcal{N}_{0,0}(n)} \right)^2 ,
\end{eqnarray}
with
\begin{equation}
\mathcal{N}_{\mu,\nu}(n)  =  {N \choose n}^{-1} \sum_{ \{ j \}_n } \left( \prod_{k \in \{ j \}_n} 1/\alpha_k^2 \right) \left[
\sum_{k \in \{ j \}_n} \alpha_k^{\mu} \right]^{\nu} . \nonumber 
\end{equation}
To estimate the required $N_{\mu,\nu}$ we average over a statistically significant fraction of the allowed $\{ j \}_n$ for
each $n$ sublevel.  
The $\alpha_k$'s are drawn from an oblate Gaussian electron wavefunction of ratio $(1,1,1/3)$, and we omit spins with $\alpha_k < 1/N$.   We plot the three parameters $\Omega_n,\chi_n,\omega_n$ versus $n$ in Figure~\ref{f:inhomog_fid}.  The perturbative treatment used above is justified as $ \Omega_n \gg \omega_n,$ and $\chi_n$ for all $n$.  It also shows that increasing $N$ improves this ratio.

\begin{figure}
~\\
~\\
\includegraphics[width=2.5 in]{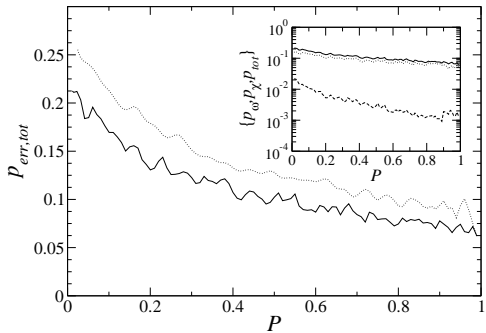} 
\caption{
Expected error of transfer and recovery for the inhomogeneous case versus final polarization $P$ for $N=4145$ (dotted) and 18924 (solid).  Iinset shows the error due to $\omega_n$ (dotted) and $\chi_n$ (dashed), and total error (solid) for $N$=18924.
\label{f:ad_fid}
}
\end{figure}

In Figure~\ref{f:ad_fid} we plot total probability of error 
for the saturated mixture as a function
of the final saturated polarization $P$.  We used an estimated hyperfine contact interaction $a N \simeq 2 \times 10^{10}~{\rm s}^{-1}$ and $\gamma \sim 6 \times 10^6~{\rm s}^{-1}$.  In all cases, adiabatic transfer requires a small change of effective field ($\simeq 100$ mT) over 10-100 ns, which could be implemented through g-factor engineering~\cite{salis} or spin-dependent optical stark shifts~\cite{imamoglu03}.  For $10^4$ nuclei, fidelities better than 0.8 are possible with realistic spin decoherence rates even with vanishingly small polarizations.  The error decreases further with increasing $N$.

In conclusion we have demonstrated that electron spin qubits can be used to effectively prepare and manipulate a local nuclear spin environment.
Specifically,
long coherence times and 
high fidelities for the storage of electron spin states into nuclear 
spins can be achieved provided the same qubit is used for the cooling process.  Such ``coherent'' cooling and storage is effective for nuclear spin preparation due to their long coherence times.  Related techniques can be used for engineering quantum states of nuclear
spins from a saturated bath state~\cite{taylor03,stateengine}.  
We further note that the techniques described in the present letter may be applicable to other  systems 
involving mesoscopic spin baths.  For example, we anticipate that similar methods may be used to prepare the local environment of superconducting qubits.

The authors would like to acknowledge helpful conversations with A. Sorensen, C.M. Marcus, and G. Giedke.  JT thanks the quantum optics group at ETH for kind hospitality during his stay.  The work at Harvard was supported by ARO, NSF, Alfred P. Sloan Foundation, and David and Lucile Packard Foundation.


\begin{thebibliography}{20}
\bibitem{decohstuff}
Even without adiabatic transfer and at low field, 
a spin-echo pulse train applied faster than $\langle \Omega_J \rangle \sim 10^8 {\rm s}^{-1}$ will entirely cancel phase and spin-exchange evolution, thereby decoupling electron spin from nuclei and completely suppressing nuclear spin-related dephasing.
\bibitem{pulsecite}
The pulse is of the form $ \Delta(t) = (\Delta_i + \Delta_f)/2 + (\Delta_i - \Delta_f)/\epsilon_1 \tan [\xi \pi (t-T/2+\epsilon_2)/2 + \epsilon_3]$
  with parameters $\epsilon_{1,2,3}$ set to give proper starting and ending detunings,
\end{thebibliography}
\end{document}